\documentclass[]{svjour3}

\RequirePackage{fix-cm}
\usepackage{amssymb}
\usepackage[T1]{fontenc}
\usepackage{textcomp}
\usepackage{amstext}
\usepackage{graphicx}
\usepackage{cite}
\usepackage{epsfig}
\usepackage{amsfonts}
\usepackage{amsmath}
\usepackage[english]{babel}
\usepackage{subfigure}
\usepackage{color}
\graphicspath{./}
\usepackage{graphicx}
\usepackage{float}
\usepackage{longtable}
\usepackage{lineno}
\smartqed  
\usepackage{graphicx}
\begin{document}

\title{Patient--specific predictions of aneurysm growth and remodeling in the ascending thoracic aorta using the homogenized constrained mixture model}
\titlerunning{Patient--specific predictions of aneurysm growth and remodeling}
\author{S. Jamaleddin Mousavi \and Solmaz Farzaneh \and St\'ephane Avril}
%
%
\institute{S. Jamaleddin Mousavi \and Solmaz Farzaneh \and S. Avril$^{*}$  \at
              Mines Saint-\'Etienne, Univ Lyon, Univ Jean Monnet, INSERM, U 1059 Sainbiose, Centre CIS, F - 42023 Saint-\'Etienne France.\\
              Phone: 0477420188, Fax: +33477420000,\\
              $^{*}$Corresponding author \email{Avril@emse.fr} }
\maketitle
\begin{abstract}
In its permanent quest of mechanobiological homeostasis, our vasculature significantly adapts across multiple length and time scales in various physiological and pathological conditions. Computational modeling of vascular growth and remodeling (G\&R) has significantly improved our insights of the mechanobiological processes of diseases such as hypertension or aneurysms. However, patient--specific computational modeling of ascending thoracic aortic aneurysm (ATAA) evolution, based on finite-element models (FEM), remains a challenging scientific problem with rare contributions, despite the major significance of this topic of research. Challenges are related to complex boundary conditions and geometries combined with layer-specific G\&R responses. To address these challenges, in the current paper, we employed the constrained mixture model (CMM) to model the arterial wall as a mixture of different constituents such as elastin, collagen fiber families and smooth muscle cells (SMCs). Implemented in Abaqus as a UMAT, this first patient--specific CMM-based FEM of G\&R in human ATAA was first validated for canonical problems such as single--layer thick--wall cylindrical and bi--layer thick--wall toric arterial geometries. Then it was used to predict ATAA evolution for a patient--specific aortic geometry, showing that the typical shape of an ATAA can be simply produced by elastin proteolysis localized in regions of deranged hemodymanics. The results indicate a transfer of stress to the adventitia by elastin loss and continuous adaptation of the stress distribution due to change of ATAA shape. Moreover, stress redistribution leads to collagen deposition where the maximum elastin mass is lost, which in turn leads to stiffening of the arterial wall. As future work, the predictions of this G\&R framework will be validated on datasets of patient--specific ATAA geometries followed up over a significant number of years.
\end{abstract}
%
%
\section*{List of symbols}
In folowing $i \in \left\lbrace {\text{e}, \text c}_j, \text{m} \right\rbrace$ and $k \in \left\lbrace {\text c}_j, \text{m} \right\rbrace$ \\
\begin{longtable}{p{2cm}p{11cm}}
${\bf a}_0^k$                                  & The unit vector pointing direction of the $k$th fiber \\
${\bf C}^i_\text{el}$                         & Elastic right Cauchy-Green deformation tensor of the $i$th constituent \\
$\overline{\bf C}^i_\text{el}$                & Modified elastic right Cauchy-Green deformation tensor of the $i$th constituent \\
$D_\text{max}$                                 & Maximum damage of elastin \\    
$\dot{D}^i_{\text g}$                          & Generic rate function of $i$th constituent \\
${\bf F}$                                      & The total deformation gradient of the mixture \\
${\bf F}^i_\text{tot}$                         & Total deformation gradient of the $i$th constituent \\
${\bf F}^i_\text{el}$                          & Elastic deformation gradient of the $i$th constituent \\
${\bf F}^i_\text{gr}$                          & Total inelastic (G\&R) deformation gradient of the $i$th constituent \\
${\bf F}^i_\text{g}$                           & Deformation gradient of the $i$th constituent due to growth \\
${\bf F}^i_\text{r}$                           & Deformation gradient of the $i$th constituent due to remodelling \\
${\bf G}^i_{\text h}$                          & Deposition stretch tensor of the $i$th constituent \\
$J$                                            & Jacobian of the mixture \\
$\overline{I}_1^i$                             & First invariant of the right Cauchy-Green deformation tensor for the $i$th constituent \\
${I}_4^i$                                      & Fourth invariant of the right Cauchy-Green deformation tensor for the $i$th constituent \\
$k^{{\text c}_j}_\sigma$                       & Gain or growth parameter of collagen fiber families \\
$k^k_1$                                        & Fung-type material coefficient the $k$th constituent \\
$k^k_2$                                        & Fung-type material coefficient the $k$th constituent \\  
$L_\text{dam}$                                 & The spatial damage spread parameter of elastin \\
${\bf S}$                                      & Second Piola--Kirchhoff stress \\
$T^i$                                          & The average turnover time of the $i$th constituent \\   
$t_\text{dam}$                                 & The temporal damage spread parameter of elastin \\
$W$                                            & The specific strain energy density function of the mixture \\
${W}^i$                                        & The strain energy of the $i$th individual constituents \\ 
${\bf X} $                                     & Material point in a reference configuration \\
${\bf x}$                                      & Material point in a deformed or current configuration \\    
$\alpha^{c_{j}}$                               & each direction of collagen fiber families \\
$\mu^{\text e}$                                & Neo Hookean material coefficient of elastin \\           
$\kappa$                                       & Bulk modulus of elastin \\           
$\sigma^i$                                     & Current stress of extant $i$th constituent \\
$\sigma^{{\text c}_j}_{\text h}$               & Average stress of $i$th constituent at homeostasis \\
$\lambda^{\text e}_z$                          & Axial elastin deposition stretch value \\ 
$\lambda^{\text e}_\theta$                     & Circumferential elastin deposition stretch value \\ 
$\lambda^k$                                    & Deposition stretch value of $k$th constituent in fiber direction \\
${\bf \Omega}_0$                               & Reference configuration \\
${\bf \Omega}(t)$                              & Deformed or current configuration \\
$\varrho^i_0$                                  & Mass densities of of the $i$th constituent before G\&R \\ 
$\varrho^i_t$                                  & Mass densities of the $i$th constituent at time $t$ \\
$\dot{\varrho}^{{\text e}}(t)$                 & The rate of mass degradation of the elastin \\
$\dot{\varrho}^{{\text c}_j}_\text{adv}(t)$    & The rate of mass degradation or deposition in the adventitia for collagen fibers \\
$\dot{\varrho}^{{\text c}_j}_\text{med}(t)$    & The rate of mass degradation or deposition in the media for collagen fibers \\
\end{longtable}
\section{Introduction}
Growth and remodeling (G\&R) are fundamental mechanobiological processes in normal tissue development and in various pathological conditions. It is suggested that G\&R in tissues may be mediated by mechanical stresses. For example, cardiac hypertrophy and normal cardiac growth develop in response to increased hemodynamic loading and altered systolic and diastolic wall stresses \cite{Grossman80}. Sustained hypertension is also associated with changes such as increased wall thickness in large arteries \cite{Humphrey08}. This adaptation ability of soft tissues is related to the existence of a mechanical homeostasis across multiple length and time scales in the vasculature. At the tissue scale, this manifests through continuous mass changes of the components of the extracellular matrix (ECM) such as collagen, elastin and proteoglycans \cite{Cyron16,Humphrey02}. 

In the current paper, we are interested in continuum finite-element formulations to simulate G\&R in arteries. The first model of mechano--regulated soft tissue growth was presented by Rodriguez et al \cite{Rodriguez94} in the mid--1990s, incorporating the associated growth by multiplicative decomposition of the total deformation gradient into an elastic and inelastic part. Thereafter, this conceptual simplicity has been widely used by others. Comellas et al \cite{Comellas16} introduced an original constitutive model to study remodeling of damaged tissue within the framework of continuum damage mechanics and open-system thermodynamics. The total damage rate was calculated as the sum of a healing rate and a mechanical damage rate. In order to couple biochemical and biomechanical damage, the healing rate was driven by mechanical stimuli and subjected to simple metabolic constraints. Although their model was based on the mixture theory, it did not account for the evolving prestretch of each constituent.

Although many theories of G\&R have modelled the tissue as a homogenized (single-constituent) solid continuum \cite{Hosseini17,Hosseini18}, the constrained mixture model (CMM) has been increasingly employed by a number of authors \cite{Baek06,Braeu17,Cyron16,Famaey18,Figueroa09,Latorre18,Lin17,Valentin12,Valentin13,Watton04,Zeinali12} to simulate G\&R in arteries, including non--homogenized \cite{Famaey18,Valentin12,Valentin13} and homogenized \cite{Braeu17,Cyron16,Lin17} approaches. For example, Valent\'in et al \cite{Valentin11,Valentin13} established a nonlinear finite element model (FEM) based on the non--homogenized constrained mixture theory (CMT) of G\&R to facilitate numerical analyses of various cases of arterial adaptation and maladaptation. Watton et al \cite{Watton04} presented the first mathematical framework to study G\&R in two-layered cylindrical membranes. In their framework the natural configurations of each individual constituent were updated at each time step and thus cell-mediated G\&R effects could be handled \cite{Humphrey12}. They introduced collagen fibre recruitment and collagen fibre density in the strain energy density functions. Baek et al \cite{Baek06} made another important contribution to study the growth of intracranial cerebral aneurysms with a constrained mixture thin-walled model, permitting to account for evolving strain energy density functions. Although both adopted a CMM (deforming different constituents altogether under a total mixture deformation gradient but having different natural reference configurations) the mathematical foundations were slightly different. The former employed a rate-based approach whilst the latter used an integral approach. The rate-based approach was shown to be more efficient and as accurate as the integral one \cite{Gasser17}. Including disease progression and evolving geometries, CMT-based models were able to predict changes in fiber orientations and quantities, degradation of elastin and loss of smooth muscle cells (SMCs). The same concept was implemented by Famaey et al \cite{Famaey18} in Abaqus \cite{Abaqus} to predict adaptation of a pulmonary autograft over an extended period. There are a number of other applied contributions considering, for example: 
\begin{itemize} 
\item two dimensional (2D) non--homogenized CMM for arterial G\&R proposed by Baek et al \cite{Baek06}, for cerebral aneurysms and extended by Valent\'in et al \cite{Valentin09a,Valentin09b,Valentin09c} for cerebral arteries 
\item the evolving geometry, structure, and mechanical properties of a representative straight cylindrical artery subjected to changes in mean blood pressure and flow in 3D \cite{Karsaj10}. 
\end{itemize}

Braeu et al \cite{Braeu17} and Cyron et al \cite{Cyron16} introduced the homogenized CMM framework for G\&R using an informal temporal averaging approach. Lin et al. \cite{Lin17} combined homogenization and the CMT to simulate the dilatation of abdominal aortic aneurysms. Their methodology is computationally less expensive than non--homogenized CMM and it can yet capture important aspects of G\&R such as mass turnover in arterial walls. Unlike non--homogenized CMMs in which one must deal with myriads of evolving configurations, the homogenized CMM is based on a single time--independent reference configuration for each material species and each point with a time--dependent inelastic local deformation of G\&R. Recently, Latorre and Humphrey \cite{Latorre18b} introduced a new rate-based CMM formulation suitable for studying mechanobiological equilibrium and stability of soft tissues exposed to transient or sustained changes, permitting direct resolution of G\&R problems with a static approach. 

Although prior work in the CMM framework have significantly improved our insights of arterial wall G\&R, they have been mostly limited to canonical problems in arterial mechanics such as 2D \cite{Baek06,Valentin09a,Valentin09b,Valentin09c} or simplified 3D cases, using membrane \cite{Cyron16} or single--layer thick--wall axisymmetric \cite{Braeu17,Lin17} approximations. Therefore, the framework still requires to be extended to more realistic and diverse analyses including patient--specific arterial geometries. To this end, several problems still need to be addressed within the CMM framework such as layer specificity, irregular boundary conditions and complex deformations. These problems can become extremely challenging in the case of ascending thoracic aortic aneurysms (ATAA) due to the simulateneous and region specific evolution of geometry, material properties \cite{Farzaneh18}, and hemodynamic loads \cite{Condemi17,Humphrey12}.

In the present work, the objective is to set up the first nonlinear FEM based on the homogenized CMT to simulate G\&R in patient--specific ATAA. After its implementation, the FEM is first validated on an idealized single--layer thick--wall cylinder. In a second stage, the model is  illustrated for a canonical problem in arterial mechanics: G\&R of a toric bilayer thick--wall arterial geometry. Then it is used to predict ATAA evolution for a patient--specific aortic geometry, showing that the typical shape of an ATAA can be obtained simply with a proteolysis of elastin localized in regions of deranged hemodymanics. 
\section{Material and methods}
\subsection{G\&R kinematics}
Let ${\boldsymbol \chi}:{\bf \Omega}_0$ be the general mapping in a ${\mathbb R}^3$ domain. ${\bf \Omega}_0$ is considered as the \textit{in vivo} (for example healthy) configuration of a blood vessel before any specific G\&R starts. The total deformation gradient of a mixture of $n$ different constituents (e.g. elastin, collagen fiber families, or SMCs), ${\bf F}$, between a material point, ${\bf X} \in {\bf \Omega}_0$, from a reference configuration, ${\bf \Omega}(0) = {\bf \Omega}_0$, and a position, ${\bf x}={\boldsymbol \chi}({\bf X},t) \in {\bf \Omega}_0$, in a deformed or current configuration, ${\bf \Omega}(t) = {\bf \Omega}_t$, at time $t$ can be defined as
\begin{equation}\label{F}
\begin{aligned}
{\bf F}({\bf X},t) = \frac{\partial {\bf x}}{\partial {\bf X}}
\end{aligned}
\end{equation}
Reference volume elements d$V \in {\bf \Omega}_0$ are mapped to volume elements d$v$ = $J$d$V \in {\bf \Omega}_t$ with the Jacobian $J = \mid {\bf F} \mid$ > 0.

According to the CMT, we assume that all constituents in the mixture deform together under the total deformation gradient ${\bf F}$ in the stressed field while each constituent has a different "total" deformation gradient resulting from its own deposition stretch. Therefore, assuming that ${\bf G}^i_{\text h}$ ($i \in \left\lbrace {\text e}, {\text c}_j, \text{m} \right\rbrace $, where superscripts e, c$_j$ and m represent respectively the elastin, the constituent made of each of the $n$ possible collagen fiber families and the SMCs, all these constituents making the mixture) is the deposition stretch of the $i$th constituent with respect to the reference homeostatic configuration  \cite{Cardamone09,Mousavi17,Mousavi18}, the "total" deformation gradient of the $i$th constituent can be calculated as
\begin{equation}\label{F_i_tot}
{\bf F}^i_\text{tot} = {\bf F} {\bf G}^i_{\text h}
\end{equation}
where the deposition stretch tensor of elastin may be written such as ${\bf G}^\text{e}_\text{h}$ = diag[$\frac{1}{\lambda^{\text e}_\theta \lambda^{\text e}_z}, \lambda^{\text e}_\theta, \lambda^{\text e}_z$] to ensure incompressibility, and the deposition stretch tensor of collagen families and SMCs may be written such as ${\bf G}^\text{k}_\text{h} = \lambda^k {\bf a}_0^k \otimes {\bf a}_0^k + \frac{1}{\sqrt{ \lambda^k}}({\bf I} - {\bf a}_0^k \otimes {\bf a}_0^k)$, $k \in \left\lbrace {\text c}_j, \text{m} \right\rbrace$. $\lambda^{\text e}_\theta$ and $\lambda^{\text e}_z$ are the deposition stretches of elastin in the circumferential and longitudinal directions, respectively, and $\lambda^k$ is the deposition stretch of the $k$th constituent in the fiber direction with a unit vector ${\bf a}_0^k$.

The local stress--free state may vary between each constituent and even between the differential mass depositions of these constituents at different time increments. Thus, for each differential mass increment of the $i$th constituent deposited at time $\tau$, the total deformation gradient of each constituent in Eq. \ref{F_i_tot} may be rewritten by a multiplicative decomposition into an elastic ${\bf F}^i_\text{el}$ and inelastic (namely G\&R) part ${\bf F}^i_\text{gr}$ as 
\begin{equation}\label{F_i_tot_decom}
{\bf F}^i_\text{tot} = {\bf F}^i_\text{el} {\bf F}^i_\text{gr}
\end{equation}
It is noteworthy that due to continuous G\&R, the inelastic G\&R deformation gradient includes the changes between the local  stress-free configurations of different mass increments resulting from deposition in a different configuration and at a different time. This process is schematically shown in Fig.~xx. Dynamic effects such as inertia or viscoelasticity can usually be neglected during G\&R as they occur at slow time scales \cite{Braeu17}.
\subsection{Mechanical Constitutive model of arterial wall}
Numerous mechanical constitutive models were introduced for arterial walls \cite{Holzapfel2000}.  In this section, a strain energy density function is defined for the different components of the arterial wall based on the CMT \cite{Braeu17,Mousavi17,Mousavi18,Wilson13,Cyron16,Cyron14}. These different components are: elastin, four different families of collagen fibers oriented in circumferential, axial, and diagonal directions and SMCs. The intima layer is disregarded here as it is very thin. Based on the mass fractions of each individual component, the specific strain energy density function may be written as \cite{Braeu17,Humphrey95,Mousavi17,Mousavi18}: 
\begin{equation}\label{SEF}
W = \varrho^{\text e}_t \big( \overline{W}^{\text e}(\overline{I}_1^{\text e}) + U(J^{\text e}_\text{el}) \big) + 
\sum_{j=1}^{n} \varrho^{{\text c}_j}_t W^{{\text c}_j}({I}_4^{{\text c}_j}) + \varrho^{\text m}_t W^{\text m}({I}_4^{\text m})
\end{equation}
where $\varrho^i_t$ and ${W}^i$ ($i \in \left\lbrace {\text c}_j, \text{m} \right\rbrace$) refer respectively to the mass densities and strain energy of the individual constituents based on the first ($\overline{I}_1^i$), fourth (${I}_4^i$) invariants and Jacobian ($J$). 

Strain energy density of elastin is described by a Neo--Hookean function in which incompressibility is enforced by a penalty function of the Jacobian \cite{Cardamone09,Mousavi17,Holzapfel2000} as
\begin{subequations}\label{SEF_elastin}
\begin{align}
& \overline{W}^{\text e}(\overline{I}_1^{\text e}) = \frac{\mu^{\text e}}{2}(\overline{I}_1^{\text e}-3)\\ 
& U^{\text e}(J^{\text e}_\text{el}) = \kappa(J^{\text e}_\text{el}-1)^2
\end{align}
\end{subequations}
where $\mu^{\text e}$ and $\kappa$ are respectively a material parameter and the bulk modulus (stress--like dimensions), and
\begin{subequations}
\begin{align}
& \overline{I}_1^{\text e} = tr(\overline{\bf C}^{\text e}_\text{el})\\
& \overline{\bf C}^{\text e}_\text{el} = {\overline{\bf F}^{\text e}_\text{el}}^{\text (T)} \overline{\bf F}^{\text e}_\text{el}\\
& \overline{\bf F}^{\text e}_\text{el} = \frac{1}{{J^{\text e}_\text{el}}^{1/3}} {\bf F}^{\text e}_\text{el}\\
& J^{\text e}_\text{el}=\text{det}({\bf F}^{\text e}_\text{el})>0
\end{align}
\end{subequations}
Noting that $\text{det} (\overline{\bf F}^{\text e}_\text{el})=1$.

The passive strain energy density of SMCs and collagen families are are described using an exponential expression respectively as \cite{Cardamone09,Mousavi17,Riveros13,Rodriguez08}
\begin{equation}\label{SEF_collagen}
W^{{\text c}_j}(I_4^{{\text c}_j}) = \frac{k_1^{{\text c}_j}}{{2k_2^{{\text c}_j}}} \left[ {\text e}^{{k_2^{{\text c}_j}} (I_4^{{\text c}_j} -1)^2} - 1 \right] 
\end{equation}
and
\begin{equation}\label{SEF_SMCs_pas}
W^{\text m}(I_4^{\text m}) = \frac{k_1^{\text m}}{{2k_2^{\text m}}} \left[ {\text e}^{{k_2^{\text m}} (I_4^{\text m} -1)^2} - 1 \right] 
\end{equation}
$k_1^{{\text c}_j}$ and $k_1^{\text m}$ are stress--like material parameters while $k_2^{{\text c}_j}$ and $k_2^{\text m}$ are dimensionless material parameters. These parameters can take different values when fibers are under compression or tension \cite{Bersi16,Mousavi18}. Noting that the fourth invariant and right Cauchy--Green stretch tensor can respectively be written such as
\begin{subequations}
\begin{align}
& I_4^k = \frac{1}{\parallel {\bf F}_{gr}^k {\bf a}_0^k \parallel^2}{\bf C}^k_\text{el} : {\bf a}^k \otimes {\bf a}^k\\
& {\bf C}^k_\text{el} = {{\bf F}^k_\text{el}}^{(T)}{\bf F}^k_\text{el}
\end{align}
\end{subequations}
where $k \in \left\lbrace {\text c}_j, \text{m} \right\rbrace$ and ${\bf a}^k = \frac{{\bf F}_{gr}^k {\bf a}_0^k}{\parallel {\bf F}_{gr}^k {\bf a}_0^k \parallel}$.

For every 3D hexahedral or tetrahedral finite element across the geometry of the artery, the same strain energy density function is assumed, however different material properties and mass densities of the individual constituents may be used for each layer (media and adventitia). 

Referring to Eq.~\ref{SEF} leads to the expression of the second Piola-Kirchhoff stress tensor as:
\begin{equation}\label{SPK}
{\bf S} = \varrho^{\text e}_t ( \overline{\bf S}^{\text e} + Jp{\bf C}^{-1} ) 
+ \sum_{i=1}^{n} \varrho^{{\text c}_j}_t {\bf S}^{{\text c}_j} +  \varrho^{\text m}_t {\bf S}^{\text m}
\end{equation}
where $\overline{\bf S}^{\text e}= 2\frac{\partial \overline{W}^{\text e}}{\partial {\bf C}}$ and ${\bf S}^j= 2\frac{\partial {W}^j}{\partial {\bf C}}$ are the second Piola-Kirchhoff stress of corresponding constituents of the mixture, ($i \in \lbrace {\text c}_j, {\text m} \rbrace$), and $p = \frac{{\text d}U^{\text e}}{{\text d}J_\text{el}}$, the hydrostatic pressure. The Cauchy stress tensor is derived from the second Piola-Kirchoff stress as
\begin{equation}\label{CS}
{\boldsymbol \sigma} = J^{-1}{\bf F}{\bf S}{\bf F}^{\text T}
\end{equation}
\subsubsection{Mass turnover and inelastic G\&R deformation gradient}
In CMT--based models, G\&R is a conceptual phenomenon during which simultaneous degradation and deposition of different constituents continuously occur. This mass turnover is a stress mediated process during which extant mass is continuously degraded and new mass is deposited into the extant matrix by a stress mediated rate \cite{Braeu17,Cyron16,Famaey18}. In this work, in two-layer arterial models, mass turnover of collagen families is mediated by SMC stresses in the media and by collagen stresses in the adventitia (for the latter, it is assumed that fibroblasts of the adventitia would be sensitive to the stresses of collagen, both in intensity and directionality). Therefore, the rate of mass degradation or deposition in the media and in the adventitia can be respectively calculated as
\begin{subequations}
\begin{align}
&\dot{\varrho}^{{\text c}_j}_\text{med}(t) = \varrho^{{\text c}_j}_t k^{{\text c}_j}_\sigma \frac{\sigma^{\text m} - \sigma^{\text m}_{\text h}}{\sigma^{\text m}_{\text h}} + \dot{D}^{{\text c}_j}_{\text g}\\
& \dot{\varrho}^{{\text c}_j}_\text{adv}(t) = \varrho^{{\text c}_j}_t k^{{\text c}_j}_\sigma \frac{\sigma^{{\text c}_j} - \sigma^{{\text c}_j}_{\text h}}{\sigma^{{\text c}_j}_{\text h}} + \dot{D}^{{\text c}_j}_{\text g} 
\end{align}
\end{subequations}
where $\varrho^{{\text c}_j}_t=\varrho^{{\text c}_j}(t)$ are mass densities of collagen families at time $t$, $k^{{\text c}_j}_\sigma$ stands for collagen growth (gain) parameter, $\sigma^{\text m}_{\text h}$ and $\sigma^{{\text c}_j}_{\text h}$ are average SMCs and collagen fiber stresses at homeostasis, $\sigma^{\text m}$ and $\sigma^{{\text c}_j}$ denote the current stress of extant collagen fibers and SMCs. 

Moreover, it is assumed that elastin can be only subjected to degradation, if any, and its mass loss cannot be compensated by new elastin deposition.
\begin{equation}\label{Elastin_loss}
\dot{\varrho}^{{\text e}}(t) = \dot{D}^{{\text e}}_{\text g}
\end{equation}
$\dot{D}^{{\text c}_j}_{\text g}$ and $\dot{D}^{{\text e}}_{\text g}$, so called generic rate function, is used to describe additional deposition or degradation due to any damage in collagen and elastin, respectively. Those are not stress mediated but can be driven by other factors like chemical degradation processes or mechanical fatigue. Besides, no mass turnover is assumed for SMCs.

Even when there is a mass balance between mass degradation and mass production ($\varrho^i_t=0$), the traction--free state changes as new mass is deposited with a prestress which is not usually identical to the current stress at which the existing mass is removed. This results in changes of the current average stress and, in turn, of the traction--free state of a constituent. Therefore, some change of the microstructure of the tissue, so-called remodeling, should accompany this mass balance. However, the local traction--free configuration of a constituent will be changed also by growth when there is no mass balance between mass degradation and production ($\varrho^i_t\neq0$). Thus in addition to this turnover--based remodeling, which is a volume preserving process, the mass turnover is generally associated with a local change of the volume by growth which accommodates the mass in a certain region of the body. Consequently, the traction--free configuration of a certain constituent should be amended by both remodeling-- and growth--related inelastic local changes of the microstructure and volume. To this end, we take advantage of the homogenized CMT--based G\&R model presented by Cyron et al \cite{Braeu17,Cyron16}. Therefore, multiplicative decomposition of the inelastic G\&R deformation gradient of the $i$th constituent deposited at different times reads
\begin{equation}\label{F_i_gr}
{\bf F}^i_\text{gr} = {\bf F}^i_\text{r}{\bf F}^i_\text{g}
\end{equation}
where ${\bf F}^i_\text{g}$ and ${\bf F}^i_\text{r}$ are inelastic deformation gradients due to G\&R, respectively. The former is related to any change in the mass per unit reference volume and the latter captures changes in the microstructure due to mass turnover. Therefore, having the net mass production rate based on \cite{Braeu17,Cyron16}, the evolution of the inelastic remodeling deformation gradient of the $i$th constituent at time $t$ is calculated by solving the following system of equations
\begin{equation}\label{F_r}
\left[\frac{\dot{\varrho}^i_t}{{\varrho}^i_t} + \frac{1}{T^i} \right] \left[ {\bf S}^i - {\bf S}^i_\text{pre} \right] =
\left[ 2\frac{\partial{\bf S}^i}{\partial{\bf C}^i_\text{el}} : ({\bf C}^i_\text{el}{\bf L}^i_\text{r}) \right] 
\end{equation}
where ${\bf S}$ is the second Piola--Kirchhoff stress and subscript "pre" denotes deposition prestress while ${\bf L}^i_\text{r} = \dot{\bf F}^i_\text{r} \, {{\bf F}^i_\text{r}}^{-1}$ is the remodeling velocity gradient. $T^i$ is the period within which a mass increment is degraded and replaced by a new mass increment, named the average turnover time.

It is assumed that elastin is not produced any longer during adulthood, it even undergoes a slow degradation with a half--life time of several decades \cite{Braeu17,Cyron16b}. Therefore, elastin growth can be basically calculated based on its degradation rate ($\dot{D}^{\text e}_{\text g}$) which in turn depends on elastin half--life time (some decades). This implies that the remodeling velocity gradient is zero, then the remodeling gradient is the identity  (${\bf L}^\text{e}_\text{r}={\bf 0}$, ${\bf \dot{F}}^\text{e}_\text{r}={\bf 0}$ and ${\bf F}^\text{e}_\text{r}={\bf I}$).

Any change in the mass of each constituent in a region of the arterial wall generally induces a local change of wall volume which can be captured by an inelastic deformation gradient namely the growth deformation gradient. The inelastic growth deformation gradient relates the change of shape and size of a differential volume element to the degraded or deposited mass in that element. Basically it is the geometrical and micromechanical features of the underlying growth process that dictates the local deformation gradient due to a certain mass change. The degradation or deposition of each constituent induces element deformations at each time increment that can be captured by an inelastic deformation gradient rate for each constituent. Based on the homogenized CMT, Braeu et al \cite{Braeu17} suggested that all constituents experience the same inelastic growth deformation gradient: ${\bf F}^i_\text{g} = {\bf F}_\text{g}$. Therefore, the total inelastic growth deformation gradient rate equals the sum of the growth--related deformation gradient rates of each individual constituent and can be obtained by
\begin{equation}\label{F_g_dot}
\dot{\bf F}_\text{g} = \sum_{i=1}^n \frac{\dot{\varrho}^i_t}{\varrho^{tot}_t \left[ {{\bf F}^i_\text{g}}^\text{-T} :{\bf a}^i_{\text g} \otimes {\bf a}^i_{\text g} \right] } {\bf a}^i_{\text g} \otimes {\bf a}^i_{\text g}
\end{equation}
where unit vector ${\bf a}^i_{\text g}$ stands for growth direction of individual constituents which, for example, can represent an anisotropic growth in arterial wall thickness direction. $\varrho^{tot}_t = \sum_{i=1}^n \varrho^i_t$ denotes total volumic mass at each time.
It is noteworthy that SMCs do not experience growth according to Eq.~\ref{F_g_dot}, which means that no mass turnover is assumed for them. However, because the second term on the left-hand side of Eq.~\ref{F_r} is never null, SMCs continuously undergo remodeling, leading to a continuous update of their reference configuration.
\subsection{Finite--Element implementation}\label{FEI}
The proposed model was implemented within the commercial FE software Abaqus \cite{Abaqus} through a coupled user material subroutine (UMAT) \cite{Farz15}. A 3D structural mesh made of hexahedral elements was reconstructed across the wall of the artery. The mesh was structural, which means that the edge of each element was locally aligned with the material directions of the artery: radial, circumferential and axial. For non--perfectly cylindrical geometries, the radial direction is defined as the outward normal direction to the luminal surface, the axial direction is defined as the direction parallel to the luminal centerline in the direction of the blood flow, and the circumferential direction is perpendicular to the two previously defined directions. It is assumed that each element is a mixture of elastin, collagen and SMCs with mass density varying regionally.

The deformation of the artery is computed for every time step corresponding to one month of real time. It is obtained by feeding equilibrium equations with the constitutive equations introduced previously, and solving the resulting nonlinear equations using the Newton Raphson method. G\& R deformations tensors are obtained at each time step based on stresses assessed at the previous step. Only the initial time step is assumed to satisfy homeostatic conditions. 
\section{Numerical Applications}
Three different models were considered: 
\begin{enumerate}
\item The first case was an thick--wall cylindrical artery responding to localized elastin loss. It was initially solved by Braeu et al. \cite{Braeu17} using the homogenized CMM and the purpose was to use these previous results for validating our model.
\item The second case was a thick--wall toric artery responding to localized elastin loss. The toric model was previously used by Alford and Taber \cite{Alford08} to study G\&R in the aortic arch.
\item The third case was a thick--wall patient--specific artery responding to localized elastin loss.
\end{enumerate}

Previous work with the homogenized CMM considered a single layer to model the artery. Similar single--layer models were used in the first case using material properties taken from \cite{Braeu17} and reported in Table~\ref{parameters}. In other cases, in order to have a more realistic model of G\&R, we considered two--layer thick--wall arteries, with different material properties for the media and the adventitia. Additional material parameters used in the two--layer thick--wall model were calibrated with data of our group \cite{Davis16,Mousavi18}, they are reported in Table~\ref{inputs_PS}. 
\subsection{Application to a single--layer thick--wall cylindrical artery responding to elastin loss}\label{cylinder}
An idealized single--layer thick--wall cylindrical artery with $r=10$~mm and $h=1.41$~mm was considered. It was assumed that this geometry, which was set as the reference configuration, was related to a reference pressure of 13.3 kPa and was at homeostasis. The deposition stretch of elastin permitting to obtain mechanical equilibrium in the reference configuration was solved using the algorithm presented in \cite{Mousavi17}. Following \cite{Braeu17}, elastin was degraded with the following rate:
\begin{equation}\label{D_elastin_cylinder}
\dot{D}^{{\text e}}_{\text g}({\bf X},t)  = -\frac{\varrho^{\text e}({\bf X},t)}{T^{\text e}} - \frac{D_\text{max}}{t_\text{dam}}\varrho^{\text e}({\bf X},0)e^{-0.5(\frac{X_3}{L_\text{dam}})^2 - \frac{t}{t_\text{dam}}}
\end{equation}
where $L_\text{dam}$ and $t_\text{dam}$ are the spatial and the temporal damage spread parameters, respectively, while $D_\text{max}$ is maximum damage. $X_3$ is the material position in the axial direction of the cylinder. Due to symmetrical geometry one--fourth of the cylinder was modeled using symmetric boundary conditions. The axial direction was defined such as $0 \leq X_3\leq \frac{L}{2}$. The first term in Eq \ref{D_elastin_cylinder} denotes a normal elastin loss by age while the second one is related to a sudden and abnormal local damage starting at $t=0$ with maximum value at the center of the cylinder ($X_3=0$) and fading at $X_3=\frac{L}{2}$. The 3D results obtained with our model on this case are compared with the corresponding 3D results of \cite{Braeu17} for six different growth parameters, $k^{{\text c}_j}_\sigma$.
\begin{table}[!hb]
\begin{center}
\caption{Material parameters employed for a single--layer thick--wall cylindrical artery and a two--layer thick--wall toric artery \cite{Braeu17}. $\alpha^{c_1}$, $\alpha^{c_2}$, $\alpha^{c_3}$ and $\alpha^{c_4}$ are axial, circumferential and two diagonal directions of collagen fiber families, respectively.}\label{parameters}
\begin{tabular}{p{5cm}l}
\hline
\textbf {Symbol}                          & \textbf {Values}\\
\hline
$\alpha^{c_{j}}, j=1,2,...,4$  & 0, $\frac{\pi}{2}$ and $\pm \frac{\pi}{4}$ \\
$\mu^{\text e}$                           & 72 [J/kg]  \\           
$\kappa$                                  & 720 [J/kg] \\           
$k_1^{{\text c}_j}$                       & 568 [J/kg] \\
$k_2^{{\text c}_j}$                       & 11.2 \\  
$k_1^{\text m}$                           & 7.6 [J/kg] \\     
$k_2^{\text m}$                           & 11.4 \\    
$\varrho^{\text e}_0$                     & 241.5 [kg/m$^3$] \\ 
$\varrho^{{\text c}_1}_0$                 & 65.1 [kg/m$^3$] \\ 
$\varrho^{{\text c}_2}_0$                 & 65.1 [kg/m$^3$] \\ 
$\varrho^{{\text c}_3}_0 = \varrho^{{\text c}_4}_0$                    & 260.4 [kg/m$^3$] \\ 
$ \varrho^{\text m}_0$                    & 157.5 [kg/m$^3$] \\ 
$\lambda^{\text e}_z$                     & 1.25  \\ 
$\lambda^{{\text c}_j}$                   & 1.062  \\ 
$\lambda^{\text m}$                       & 1.1 \\ 
$T^{\text e}$                             & 101 [years] \\   
$T^{{\text c}_j}$                         & 101 [days] \\   
$T^{\text m}$                             & 101 [days] \\   
$L_\text{dam}$                            & 10 [mm] \\  
$t_\text{dam}$                            & 40 [days] \\
$D_\text{max}$                            & 0.5 \\                      
\hline
\end{tabular}
\end{center}
\end{table}
\subsection{Application to a two--layer thick--wall toric artery responding to elastin loss}\label{arch}
We employed the model on a torus shown in Fig.~xx with $\frac{R}{r}=4$. Its thickness and inner radius were assumed identical to the ones of the cylindrical artery defined in section~\ref{cylinder} ($r=10$~mm and $h=1.41$~mm). Due to symmetry only a quarter of the torus was modeled using symmetric boundary conditions.

Again, it was assumed that this geometry, which was set as the reference configuration, was related to a reference pressure of 13.3 kPa and was at homeostasis. The deposition stretch of elastin permitting to obtain mechanical equilibrium in the reference configuration was solved using the algorithm presented in \cite{Mousavi17,Maes19}. 

Then, an elastin degradation rate with temporal and spatial damage was assumed as
\begin{equation}\label{D_elastin_arch}
\dot{D}^{{\text e}}_{\text g}({\bf X},t)  = -\frac{\varrho^{\text e}({\bf X},t)}{T^{\text e}} - \frac{D_\text{max}}{t_\text{dam}}\varrho^{\text e}({\bf X},0)e^{-0.5(\frac{\theta}{\theta_\text{dam}})^2 - \frac{t}{t_\text{dam}}}
\end{equation}
where $\theta_\text{dam}$ is the spatial damage spread parameters. $0 \leq \theta \leq 90$ is the material position varying as shown in Fig.~xx, indicating maximal and minimal elastin loss at $\theta = 0^\circ$ and $\theta = 90^\circ$, respectively. 

Material parameters reported in Table~\ref{parameters} were employed considering that the media comprised 97\% of total elastin, 100\% of total SMCs, and 15\% of total axial and diagonal collagen fibers while the adventitia comprised 3\% of total elastin, 85\% of total axial and diagonal collagen, and 100\% of total circumferential collagen \cite{Bellini14}.
\subsection{Application to a two--layer thick--wall patient--specific human ATAA responding to elastin loss}\label{ATAA}
To demonstrate the applicability of the model to predict patient--specific wall G\&R, the model was employed onto the geometry of a real human ATAA. An ATAA specimen and the preoperative CT scan of the patient were obtained after informed consent from a donor undergoing elective surgery for ATAA repair at CHU--SE (Saint-Etienne, France). The lumen of the aneurysm was clearly visible in the DICOM file, but detection of the aneurysm surface was not possible automatically. A non--automatic segmentation of the CT image slices was performed using MIMICS (v. 10.01, Materialise NV) to reconstruct the ATAA geometry. The reconstructed geometry was meshed with 7700 hexahedral elements. A wall thickness of 2.38 mm was defined evenly in the reference configuration, yielding an average thickness of 2.67 mm at zero pressure, which corresponded to the measured thickness on the supplied sample \cite{Farzaneh18}. Material parameters (reported in Table~\ref{inputs_PS}) such as deposition stretch of collagen and exponents were taken from literature \cite{Bellini14,Cardamone09} and others were calibrated with data of our group \cite{Davis16}. Note that 97\% of total elastin, 100\% of total SMC, and 15\% of total axial and diagonal collagen fibers were assigned to the media. Conversely, 3\% of total elastin, 85\% of total axial and diagonal collagen, and 100\% of total circumferential collagen were assigned to the adventitia \cite{Bellini14,Mousavi17,Mousavi18}. The geometry obtained from the CT scan was assigned as the reference configuration. It was subjected to a luminal pressure of 80~mmHg (diastole). An axial deposition stretch of 1.3 was defined for the elastin and the deposition stretches of collagen and SMC components were set to 1.1. The spatially varying circumferential deposition stretch of elastin was determined to ensure equilibrium with the luminal pressure using our iterative algorithm \cite{Mousavi17}.  Both ends of the ATAA model were fixed in axial and circumferential directions, allowing only radial displacements.

4D flow magnetic resonance imaging (MRI) with full volumetric coverage of ATAAs can reveal complex aortic 3D blood flow patterns, such as flow jets, vortices, and helical flow \cite{Condemi17,Hope07}. For the same patient, 4D flow MRI datasets were also acquired, revealing a jet flow impingement against the aortic wall around the bulge region (downstream the area of maximum dilatation) as shown in Fig~xx. Guzzardi et al. \cite{Guzzardi15} found that regions with largest WSS underwent greater elastin degradation associated with vessel wall remodeling in comparison with the adjacent regions with normal WSS. Consequently, based on these findings we considered a localized elastin degradation and simulated its effects on ATAA G\&R. The local elastin degradation rate was written such as
\begin{equation}\label{D_elastin_PS}
\dot{D}^{{\text e}}_{\text g}({\bf X},t)  = -\frac{\varrho^{\text e}({\bf X},t)}{T^{\text e}} - \frac{D_\text{max}}{t_\text{dam}}\varrho^{\text e}({\bf X},0)e^{-\frac{t}{t_\text{dam}}}
\end{equation}
Three different values of $t_\text{dam}$ (as listed in Table~\ref{inputs_PS}) were studied.
\begin{table}[!hb]
\begin{center}
\caption{Material parameters employed for two--layer patient--specific human ATAA model adapted from \cite{Mousavi18}. $\alpha^{c_1}$, $\alpha^{c_2}$, $\alpha^{c_3}$ and $\alpha^{c_4}$ are axial, circumferential and two diagonal directions of collagen fiber families, respectively.}\label{inputs_PS}
\begin{tabular}{p{5cm}l}
\hline
\textbf {Symbol}                          & \textbf {Values}\\
\hline
$\alpha^{c_{j}}, j=1,2,...,4$  & 0, $\frac{\pi}{2}$ and $\pm \frac{\pi}{4}$ \\
$\mu^{\text e}$                                        & 82 [J/kg]  \\           
$\kappa$                                               & 100$\mu^{\text e}$ [J/kg] \\           
$k_1^{{\text c}_j,{\text c}}=k_1^\text{m,c}$           & 15 [J/kg] \\
$k_2^{{\text c}_j,{\text c}}=k_2^\text{m,c}$           & 1.0 \\  
$k_1^{{\text c}_j,{\text t}}$                          & 105 [J/kg] \\
$k_2^{{\text c}_j,{\text t}}$                          & 0.13 \\  
$k_1^\text{m,t}$                                       & 10 [J/kg] \\     
$k_2^\text{m,t}$                                       & 0.1 \\    
$\varrho^{\text e}_0$                                  & 250 [kg/m$^3$] \\ 
$\varrho^{{\text c}_j}_0$                              & 460 [kg/m$^3$] \\ 
$ \varrho^{\text m}_0$                                 & 280 [kg/m$^3$] \\ 
$\lambda^{\text e}_z$                                  & 1.3  \\ 
$\lambda^{{\text c}_j}$                                & 1.1  \\ 
$\lambda^{\text m}$                                    & 1.1 \\ 
$T^{\text e}$                                          & 101 [years] \\   
$T^{{\text c}_j}$                                      & 101 [days] \\   
$T^{\text m}$                                          & 101 [days] \\   
$t_\text{dam}$                                         & 20, 40 and 80 [days] \\
$D_\text{max}$                                         & 0.5 \\                      
\hline
\end{tabular}
\end{center}
\end{table}
\section{Results}
\subsection{Response of a single--layer thick--wall cylindrical artery to localized elastin loss}\label{R-cylinder}
The dilatation at the middle of the cylindrical artery (where maximum elastin loss occurs) is shown in Fig.~xx over 15 years for different growth parameters. There are a good agreement between these results obtained with our 3D FEM and the numerical model from \cite{Braeu17}. The cylindrical artery responds to elastin degradation with large and unstable dilatations for small gain parameters while it only slightly dilates in the case of relatively large gain parameters, recovering its stability after a transient period. Distribution of maximum principal stresses and distributions of normalized collagen mass density for the largest and smallest gain parameters are shown in Figs.~xx. Elastin loss naturally leads to higher stresses in the other components of the arterial wall and subsequently higher deposition of new collagen fibers.
\subsection{Response of a two--layer thick--wall toric artery to localized elastin loss}\label{R-arch}%
The effect of elastin loss during 15 years in a two--layer toric artery was considered for $k^{{\text c}_j}_\sigma = \frac{0.05}{T^{{\text c}_j}}$ and $k^{{\text c}_j}_\sigma = \frac{0.15}{T^{{\text c}_j}}$. The change of the thickness and diameter due to degradation of the elastin are shown in Figs.~xx and Figs.~xx, respectively. The dilatation and thickness were never stable for small growth parameters ($k^{{\text c}_j}_\sigma = \frac{0.05}{T^{{\text c}_j}}$). Conversely, for relatively large growth parameters ($k^{{\text c}_j}_\sigma = \frac{0.15}{T^{{\text c}_j}}$), the thickness and diameter became stable after about five years of transient growth period. The wall was basically thickened on the outer curvature side, mainly in the media. Therefore, the response of a toric artery to elastin loss is unstable for small growth parameters while it recovers its stability, after some enlargement, for relatively large growth parameters. In addition, colormaps of the maximum principal stress and the collagen mass density for large and small growth parameters (Figs.~xx) show that elastin loss continuously causes higher stresses and collagen deposition in the media. However, the balance between arterial dilatation and collagen deposition leads to higher collagen production for small gain parameters. This in turn ends with higher stresses in the arch with small gain parameters.
\subsection{Response of a two--layer patient--specific human ATAA to localized elastin loss}\label{R-ATAA}
The G\&R response of a patient--specific ATAA to localized elastin degradation is shown in Fig.~xx-a, c and e. Due to change of shape, the stress distribution is in continuous adaptation. For all cases, elastin loss induces a transfer of stress to the adventitia in the damaged region. Moreover, an increase of $t_\text{dam}$ results in an increase of maximum principal stresses in the arterial wall. It is induced by the related increase of elastin degradation rate. In Fig.~xx-b, d and f, the distribution of collagen mass density for different $t_\text{dam}$ shows that most of the collagen is deposited in the media where elastin has been lost (recall that $\sim$97\% of the elastin is in media), causing finally a thickening of the arterial wall (Fig.~xx). It is noteworthy that increase of $t_\text{dam}$ accelerates collagen deposition and consequently wall thickening. Moreover, we studied the sensitivity of ATA dilatation to the collagen growth parameter, $k^{{\text c}_j}_\sigma$. As shown in Fig.~xx, larger growth parameters stabilize ATA dilatation induced by elastin loss. However, for relatively small growth parameters, $k^{{\text c}_j}_\sigma = \frac{0.1}{T^{{\text c}_j}}$, as shown in Figs.~xx-a, xx-b and xx-a, the ATA undergoes an excessive dilatation (the maximum ATA diameter increases continuously from  $\sim$42 mm to  $\sim$64 mm after $\sim$180 months). As the newly deposited collagen has to compensate for the elastin loss to maintain the homeostatic state, this induces the adaptation response. Conversely, increasing the growth parameter leads to a stable growth of ATA after 31 months (the maximum diameter of ATA after elastin loss stops increasing after $\sim$31 months, enlarging from  $\sim$42 mm to  $\sim$47 mm). For all cases, whatever the growth parameter, remodeling induced by collagen deposition always causes ATA wall thickening, mainly in the media (see Fig.~xx).
\section{Discussion}
A robust computational model based on the homogenized CMT was presented and its potential was shown to predict ATAA evolution for a patient--specific aortic geometry, showing that the typical shape of an ATAA can be obtained simply with a proteolysis of elastin localized in regions of deranged hemodymanics. The most interesting result is that although elastin degradation occurs locally in the ATAA at the location of WSS peak, the whole ATAA globally undergoes G\&R due to redistribution of stresses distribution, leading to ATAA dilatation.

A general advantage of the model presented here is that it was developed to account for \textit{in situ} prestrain (and therefore prestress). It permits to run FE analysis of G\&R in soft biological tissues without requiring a zero--pressure configuration. This appears to be especially beneficial when a patient--specific geometry is reconstructed using CT scans or MRI data acquired in a pressurized configuration. Using this methodology, the prestress is calculated based on the prestrain, defined in terms of fiber prestretches (deposition stretches), assuming a hyperelastic elastic material behavior. Therefore, a drawback of this methodology is that we may encounter some instability in the resolution if we enforce a particular deposition stretch for each constituent. Large distortions of elements may also occur with small variations of deposition stretchs and lead to the divergence of the resolution. This indicates that an arbitrary deposition stretch cannot be always imposed on an arbitrary reference configuration.

For all geometries given herein for large gain parameters, $k_\sigma$, the blood vessel recovered a new stable state after a transient period of dilation and enlargement. In contrast, for small gain parameters it underwent unbounded dilation, experiencing mechanobiological instability. Dilatation due to weakening of the arterial wall by elastin loss is physically consistent with previous findings \cite{Braeu17,Cyron16}. However, one can find 3D FE implementation of G\&R in which the arterial radius decreases after elastin degradation \cite{Eriksson14,Valentin13,Grytsan15}. This can be explained by the implementation of volumetric growth. \cite{Eriksson14,Valentin13,Grytsan15} defined implicitly the growth directions using the volumetric deviatoric contributions of the deformation gradient and imposing incompressibility constraints. Only isotropic growth can be modeled with their approach, elastin degradation consequently causing a decrease of total tissue volume. Therefore, in their model the arterial wall shrinks in all spatial directions, including the circumferential direction. Eriksson \cite{Eriksson14b} attempted to overcome this problem by introducing the concept of constant and adaptive individual density growth in which an elastin loss does not cause a contraction of the arterial wall due to loss of mass. Nevertheless, it is still controversial whether elastin loss would locally lead to arterial shrinking in the thickness direction or if it would really induce a change in the mass density of the tissue.

Basically, two major approaches were so far proposed for numerical modeling of soft tissue G\&R. Rodriguez et al. \cite{Rodriguez94} introduced a kinematic growth theory by multiplicative decomposition of the deformation gradient into an elastic and an inelastic growth contributions. The elastic part ensured geometric compatibility and mechanical equilibrium while the inelastic growth part contained the local changes of mass and volume. Although their model was computationally efficient and conceptually simple it was intrinsically unable to compute the separate G\&R of structurally different constituents. This limitation was fixed by CMT--based model introduced by \cite{Humphrey02} in which the \textit{in vivo} situation can be realistically mimicked using the concept of deposition stretches. The computational cost of the classical CMT--based models is higher than that of simple kinematic growth theory. Homogenized CMT--based models introduced by Cyron et al. \cite{Cyron16} combines the advantages of both classical CMT--based models and simple kinematic growth models to overcome the drawback of each model. The results obtained by homogenized CMT--based models are similar to the classical ones but with low computational cost. Focusing on the example of simple membrane--like \cite{Cyron16} and thick--wall \cite{Braeu17} cylindrical vessels, they showed that homogenized CMT--based models are able to reproduce realistically both pathological growth responding to an elastin loss (as observed in aneurysms) and adaptive growth in healthy vessels due to hypertension. The prominent privilege of CMT--based models is the inherent incorporation of anisotropic volumetric growth in the thickness direction of arterial wall (proved by experimental observations of \cite{Matsumoto96}). Moreover, recently Lin et al. \cite{Lin17} combined homogenization and CMT to capture G\&R of the abdominal aorta and to consider dilatation of abdominal aortic aneurysms under loading. They focused on a transversely isotropic mixture subjected to uniform extension in the direction of collagen fibers assuming that they are embedded in an isotropic elastin matrix, ignoring the role of SMCs. Considering a very special case of isotropic growth, their model can successfully predict the continuous enlargement of an abdominal aortic aneurysm by combined effects of elastin degradation, loss of extant collagen and production of new collagen, as well as fiber remodeling.

As anisotropic growth may stabilize the arterial wall under perturbations more efficiently than the isotropic growth \cite{Braeu17} so that the ability of the homogenized CMT--based model implemented herein can be considered an ideal tool to realistically study the patient--specific geometries undergoing G\&R in response to an unexpected degradation of elastin. Following \cite{Braeu17,Watton04,Watton09,Bellini14,Cardamone09}, it was assumed that patient--specific G\&R resulted from specific temporal and spatial distributions of elastin degradation. We considered multiple temporal damage parameters for elastin degradation leading to different aneurysm growth rate. Although the global shape of the aneurysm resembles, the thickening and collagen production rates are different for different cases. Different temporal damage constants showed significant effects on the expansion rate where the higher $t_\text{dam}$ delivers the higher G\&R rate. Although we simply employed temporal damage parameter for elastin degradation, elastin degradation during ATAA growth involves multiple biological and mechanical parameters including abnormal distribution of wall shear stress \cite{Guzzardi15} and circumferential stress \cite{Humphrey02}. The formation of intraluminal thrombus is specific to AAA \cite{Vorp01,Vorp05}. It may stimulate proteolytic effects but this was not considered here as thrombus are very rare in ATAAs. 

In the patient--specific study  it also appears that collagen deposition tends to compensate the elastin loss. It is worth noting that as aneurysm grows, principal stress may not increase necessarily in a damaged location. This is observed in AAA growth as well \cite{Zeinali11}. Moreover, although elastin was degraded locally, dilatation of the ATA was spread across a larger area due to stress redistribution.

The \textit{in vivo} images were obtained when the artery was under pressure so that the stress--free or zero pressure configuration was not basically available. Hence, for hyperelastic models such as Holzapfel--type models \cite{Holzapfel2000}, approaches such as inverse elastostatic methods \cite{Zhou10} or Lagrangian-Eulerian formulations \cite{Gee10} are required to estimate the stress--free state of \textit{in vivo} geometries obtained from medical images. One of the advantages of current CMT-based model is that G\&R analysis of a patient-specific model can be directly performed on the \textit{in vivo} geometry reconstructed from medical images obtained under pressure, without needing to compute the stress--free geometry.

Salient features of the response of arterial walls to altered hemodynamics \cite{Baek06b,Baek07,Valentin09a,Cardamone10} were captured by 2D and 3D CMT-based models \cite{Humphrey02,Cardamone12,Karsaj10}. Despite the major interest of this prior work on CMT-based models, two novelties can be highlighted in our work: application of CMT-based models to patient-specific geometries and integration of layer-specific properties (media and adventitia). Future work will focus on fully coupling the present model with CFD analyses \cite{Condemi17} to study the effects on aortic G\&R of different hemodynamic metrics, such as helicity, wall shear stress (WSS), time averaged WSS (TAWSS), oscillatory shear index (OSI) or relative residence time (RRT). Such fluid-solid-growth simulations have already been developed by different authors for cerebral \cite{Watton09b} or abdominal \cite{Sheidaei11,Grytsan15,Marsden15,Achille17} aneurysms and we will extend them to ATAA to provide additional insight into the evolution of these aneurysms.

This altogether indicates that the present model has the potential for clinical applications to predict G\&R of patient--specific geometries if a realistic rate of elastin loss and collagen growth parameter are available.\\

There are still several limitations and technical challenges associated with current model:
\begin{itemize}
\item The active role of SMCs is not considered in the present model, despite its major role in mechanosensing \cite{Humphrey15}. 

\item Theory of G\&R is based on a key assumption, the existence of mechanical homeostasis \cite{Humphrey08b,Kassab08}. It is difficult to have the assumption of a homeostatic state satisfied at every point of the arterial wall. For an idealized model, such as ideal thick--wall cylinders, the \textit{in vivo} material properties are typically assumed to be uniform across the domain. When a patient--specific geometry is used for a clinical study, it will be essential to prescribe the distribution of material and structural parameters such as thickness and fiber orientations consistent with \textit{in vivo} data. Therefore, considering the arterial wall with a uniform thickness can be considered as additional limitations of the current work.

\item Another difficulty associated with patient--specific models is estimating the constitutive parameters of the model for different patients. Here, these parameters were estimated by curve fitting from the \textit{ex vivo} bulge inflation data of an ATAA segment excised after the surgical intervention of the same patient. However, in clinical applications, it will be needed to identifying noninvasively the \textit{in vivo} material properties of ATAAs \cite{Farzaneh18,Farzaneh19}.

\end{itemize}
\section{Conclusion}
In summary, in this manuscript, a robust computational model based on the homogenized CMT was presented and its potential was shown for patient--specific predictions of growth and remodeling of aneurysmal human aortas in response to localized elastin loss. As future application, the predictions of this G\&R framework will be validated on datasets of patient--specific ATAA geometries followed up over a significant number of years.
\section{Acknowledgements}
The authors are grateful to the European Research Council for grant ERC-2014-CoG BIOLOCHANICS. The authors would also like to thank Nele Famaey (KU Leuven, Belgium), Christian J. Cyron (TU Hamburg, Germany) and Fabian A. Braeu (TU M\"{u}nchen, Germany) for inspiring discussions related to this work.
\section{Conflict of interest}
There is no conflict of interest.
%

%
\clearpage
\end{document}